\documentclass[conference]{IEEEtran}
\IEEEoverridecommandlockouts

\usepackage{cite}
\usepackage{amsmath,amssymb,amsfonts}
\usepackage{algorithmic}
\usepackage{graphicx}
\usepackage{textcomp}
\usepackage{xcolor}
\usepackage{float}
\usepackage{subfigure,hyperref}
\hypersetup{
	colorlinks=true,
	linkcolor=blue,
	filecolor=blue,      
	urlcolor=black,
	citecolor=blue,
}

\newcommand\blfootnote[1]{%
	\begingroup
	\renewcommand\thefootnote{}\footnote{#1}%
	\addtocounter{footnote}{-1}%
	\endgroup
}

\usepackage{array}
\newcommand{\PreserveBackslash}[1]{\let\temp=\\#1\let\\=\temp}
\newcolumntype{C}[1]{>{\PreserveBackslash\centering}p{#1}}
\newcolumntype{R}[1]{>{\PreserveBackslash\raggedleft}p{#1}}
\newcolumntype{L}[1]{>{\PreserveBackslash\raggedright}p{#1}}

\def\BibTeX{{\rm B\kern-.05em{\sc i\kern-.025em b}\kern-.08em
    T\kern-.1667em\lower.7ex\hbox{E}\kern-.125emX}}
\begin{document}

\title{DARNet: Dual-Attention Residual Network for Automatic Diagnosis of COVID-19 via CT Images}

%

\author{\IEEEauthorblockN{Jun Shi\IEEEauthorrefmark{1}, 
		Huite Yi\IEEEauthorrefmark{1},  
		Shulan Ruan\IEEEauthorrefmark{1}, 
		Zhaohui Wang\IEEEauthorrefmark{1},
		Xiaoyu Hao\IEEEauthorrefmark{1},
		Hong An\IEEEauthorrefmark{1} and 
		Wei Wei\IEEEauthorrefmark{2}}
		\IEEEauthorblockA{\IEEEauthorrefmark{1}School of Computer Science and Technology, University of Science and Technology of China, Hefei, China}
		\IEEEauthorblockA{\IEEEauthorrefmark{2}Department of Radiology, The First Affiliated Hospital, Division of Life Sciences and Medicine, \\ University of Science and Technology of China, Hefei, China} 
}


\maketitle

\begin{abstract}
The ongoing global pandemic of Coronavirus Disease 2019 (COVID-19) poses a serious threat to public health and the economy. Rapid and accurate diagnosis of COVID-19 is crucial to prevent the further spread of the disease and reduce its mortality. Chest Computed tomography (CT) is an effective tool for the early diagnosis of lung diseases including pneumonia. However, detecting COVID-19 from CT is demanding and prone to human errors as some early-stage patients may have negative findings on images. Recently, many deep learning methods have achieved impressive performance in this regard. Despite their effectiveness, most of these methods underestimate the rich spatial information preserved in the 3D structure or suffer from the propagation of errors. To address this problem, we propose a Dual-Attention Residual Network (DARNet) to automatically identify COVID-19 from other common pneumonia (CP) and healthy people using 3D chest CT images. Specifically, we design a dual-attention module consisting of channel-wise attention and depth-wise attention mechanisms. The former is utilized to enhance channel independence, while the latter is developed to recalibrate the depth-level features. Then, we integrate them in a unified manner to extract and refine the features at different levels to further improve the diagnostic performance. We evaluate DARNet on a large public CT dataset and obtain superior performance. Besides, the ablation study and visualization analysis prove the effectiveness and interpretability of the proposed method.  


\end{abstract}


\section{Introduction}

\blfootnote{The work is supported by the National Key Research and Development Program of China(GrantsNo.  2017YFB0202002).}

The Coronavirus Disease 2019 (COVID-19), caused by the severe acute respiratory symptom coronavirus 2 (SARS-CoV-2), is spreading rapidly across the world through extensive person-to-person transmission \cite{1}. The World Health Organization (WHO) officially declared the COVID-19 a pandemic on 11 March 2020. As of 23 August 2021, the COVID-19 has infected more than 211 million people in more than 192 countries and territories and caused more than 4.43 million deaths \cite{2}. Due to the high infectivity and fatality rate, the COVID-19 pandemic has had a devastating impact on public health and the economy. It is of great importance to conduct early diagnosis of COVID-19, for preventing the further spread of the disease and delivering proper treatment regimen. The real-time reverse transcription-polymerase chain reaction (RT-PCR) test is the golden standard for the diagnosis of COVID-19 infection \cite{3}. However, the high false-negative rate \cite{1} of RT-PCR may delay the diagnosis of potential cases. 


As a complementary strategy, Chest X-ray and Computed Tomography (CT) are widely used in the early diagnosis of patients suspected of SARS-CoV-2 infection \cite{6}. Compared with X-ray images, chest CT scans have higher sensitivity in diagnosing COVID-19 infection, and can provide more detailed information about the lesion, which is helpful for quantitative analysis \cite{7}. Early investigations have observed typical radiographic features on chest CT images such as ground-glass opacities (GGO), multifocal patchy consolidation, and vascular dilation in the lesions \cite{8,9,10,11}. However, detecting COVID-19 from CT images is demanding and prone to human errors as some early-stage patients may have normal imaging features. Besides, the similar imaging findings between COVID-19 cases and common pneumonia (CP) cases on the image make it difficult to differentiate. 

Recently, many deep learning methods have been applied to the automatic diagnosis of COVID-19 using chest CT images and achieved impressive performance. Some keyframe-based methods \cite{10,20} use local abnormal slices rather than 3D images to make diagnostic decisions, while \cite{21,22,23} focus on segmenting the lesion area and then extract specific features for diagnosis. Despite their effectiveness, most of these methods provide a multi-phase framework, which means that the errors in upstream tasks will propagate backwards. For instance, the keyframe-based methods highly rely on the accurate classification of abnormal slices, otherwise incorrect results will negatively affect subsequent tasks. Furthermore, these methods usually have high requirements for annotation data because of the additional upstream tasks. Based on traditional 2D neural networks, other methods \cite{36,39} make efforts on extending them to classify 3D CT images and obtain promising results. However, the simple network transformation has limitations in taking full advantage of the 3D properties of CT images, resulting in the diagnostic performance that may not meet actual clinical needs.

To this end, in this paper, we propose a dual-attention residual network (DARNet), to automatically diagnose COVID-19 from CP and healthy people using CT images. In DARNet, the 3D variant of ResNet-18 \cite{13} is used as the backbone network, which takes a full 3D chest CT image as input. To fully leverage the 3D spatial information, we design a dual-attention module to extract and refine the representation features at different levels. The module mainly consists of two parts: 1) channel-wise attention and 2) depth-wise attention. The former is first proposed in \cite{25}, and we implement its 3D extension. In this study, we develop the latter, which can adaptively assign depth-level weights to each feature map during the training. We evaluate our method on the largest public CT image dataset, to the best of our knowledge. The experimental results show that DARNet is superior to existing methods. We further provide ablation studies and prove the effectiveness of the proposed dual-attention module in improving the classification accuracy and the interpretability of the model.

As a summary, our work has three major contributions as follows: 

\begin{itemize}
	\item We propose DARNet to realize automatic and accurate diagnosis of COVID-19 using 3D chest CT images. In addition to superior classification performance, our method is more sensitive to the location of the lesion regions in visual attention.
	\item To make full use of 3D spatial information of CT images, we design a dual-attention module, which can refine the learned features at different levels. The experimental results prove the effectiveness of this module in improving the classification performance and the interpretability.
	\item We evaluate DARNet on a large public dataset, achieving accuracy of 93.28\%, sensitivity of 96.86\%, specificity of 97.19\%, F1-score of 95.49\%, and an area under the receiver operating characteristic curve (AUC) of 0.995.
\end{itemize}

\section{Related Work}

\subsection{Automatic Diagnosis of COVID-19}

Recently, the successful application of artificial intelligence (AI)  in medical image analysis \cite{17} has promoted the development of radiological diagnosis technology. To combat the current pandemic, plenty of research efforts had been carried out over the past few months to design an AI system for the early diagnosis of COVID-19 via radiological imaging. \cite{19,40,41} employed convolutional neural networks (CNNs) to automatically identify COVID-19 infection from chest X-ray images and obtained impressive results. However, these methods are still limited due to the low contrast and the lack of significant features caused by the high overlapping of ribs and soft tissues. 

Compared with a single X-ray image, a chest CT scan composed of hundreds of 2D slices can reflect more detailed radiographic features about the lesions, such as GGO and consolidation. To simplify the computation, several keyframe-based methods \cite{10,20} were proposed to diagnose COVID-19 in CT images and achieved promising results. But these methods underestimated the 3D spatial information of CT images and highly relied on the accurate detection of abnormal slices. \cite{21,22,23} proposed the segmentation-based approaches that can generate more specific lesion information, such as the number and volume of lesions, which was valuable for the quantitative analysis in COVID-19 diagnosis. However, obtaining large amounts of CT data with segmentation labels is the primary challenge of these methods. Besides, most of the above methods provide a multi-stage framework, which means that these methods may be affected by error propagation. \cite{36,39} directly transfer 2D neural networks to classify 3D CT images, but their performance may not meet actual clinical needs. We thus develop DARNet to diagnose COVID-19 in an end-to-end fashion, which takes a complete chest CT image as input and can achieve competitive classification performance.

\subsection{Attention Mechanism}
Attention mechanism is an effective way to improve network performance by enhancing the learned features. Hu et al. \cite{25} proposed the channel-wise attention (CA) to refine the hidden features in the channel level during training, which can make the network more focused on the important regions. In other words, the CA module amplifies the difference between channel features by highlighting the features with a greater response, and suppressing the others. The most important is that this adjustment mechanism is completely dynamic and learnable. The effectiveness of the CA module has been proved in many applications \cite{27,28,29}. At the same time, there have been many variations and extensions. For example, \cite{42,43} proposed a joint attention module based on the CA module, which brings a significant improvement in segmentation performance. These studies show that multi-attention fusion has great potential in improving network performance. Inspired by this, we design a novel attention mechanism called depth-wise attention (DA) to recalibrate the depth-level features. By combining this module with the CA module, we construct a dual-attention module to improve the representation ability of the 3D neural networks.


\section{Methods}


\begin{figure*}[htbp]
	\centering
	\subfigure[The overall architecture of DARNet.]{
		\centering
		\includegraphics[width=1.6\columnwidth]{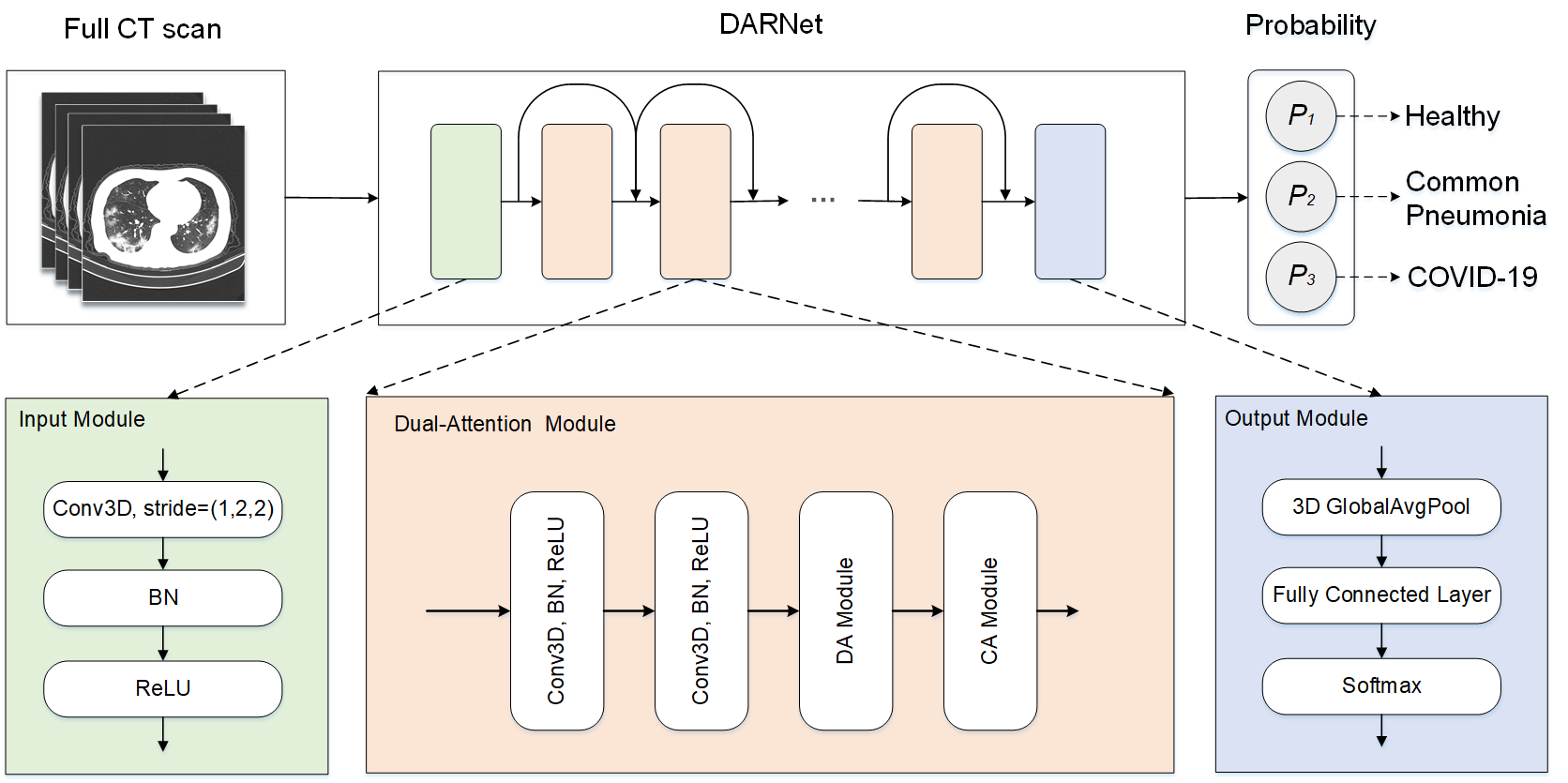}
		\label{net} 
	}
	\quad
	\subfigure[The schema of 3D CA module.]{
		\centering
		\includegraphics[width=0.8\columnwidth]{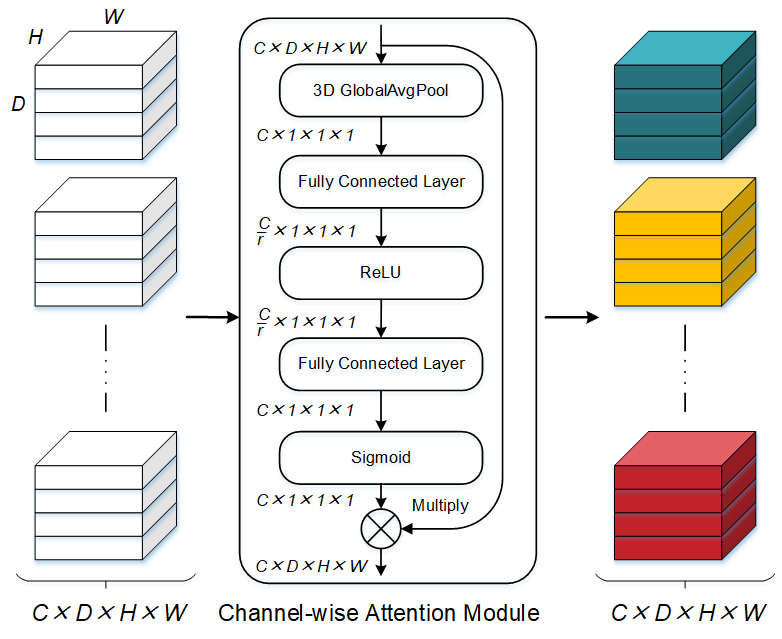}
		\label{ca} 
	}
	\quad
	\subfigure[The schema of 3D DA module.]{
		\centering
		\includegraphics[width=0.8\columnwidth]{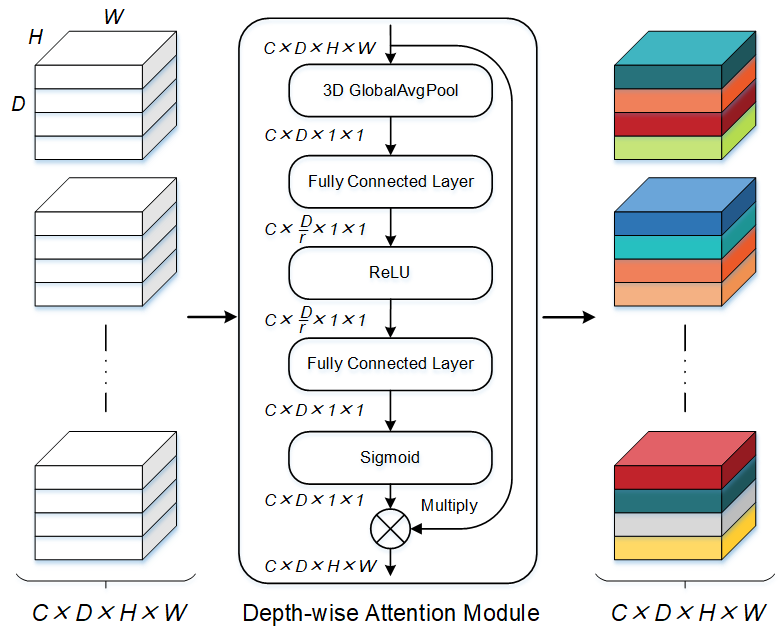} 
		\label{da}
	}
	
	\caption{Illustration of the DARNet, channel-wise attention (CA) and depth-attention (DA) modules in our method. $D$, $H$, $W$, and $C$ represent the depth, height, width, and input channels of the feature map, and $r$ refers to the reduction ratio.}
	\label{method}
\end{figure*}

\subsection{Overall Architecture}

As shown in Fig. \ref{net}, the overall architecture of DARNet mainly consists of three submodules: 1) input module, 2) dual-attention module, and 3) output module. Considering the computation complexity and GPU memory capacity, we use the 3D ResNet-18 \cite{13} as the backbone network. Specifically, the input module is composed of a 3D convolutional layer (Conv3D) with a kernel size of $(3,7,7)$ and a stride of $(1,2,2)$, a batch normalization layer (BN), and a ReLU activation layer. Besides, unlike naive ResNet-18, we remove the max-pooling layer. In this way, the input 3D CT image is downsampled by a factor of 8 in the depth dimension and a factor of 16 in the other two dimensions. The higher-resolution feature maps retain more contextual information, which is also conducive to visual analysis. In the feature extraction part, a total of 8 dual-attention modules with residual connections constitute the main structure. Each dual-attention module consists of two consecutive convolutional layers with a kernel size of $(3,3,3)$, followed by BN, ReLU, and two attention mechanisms: 1) channel-wise attention and 2) depth-wise attention. More detailed information about this module is introduced in the next subsection. For the output module, the global average pooling layer (GAP) is first used to squeeze the input features. Then a followed fully connected layer with a softmax layer generates corresponding prediction probabilities. Finally, the network returns the predicted category based on the probabilities.

\subsection{Dual-Attention Module}

A complete CT image is usually composed of hundreds of 2D slices stacked in sequence. These slices have high spatial continuity and content relevance, constituting the complete contextual information of the lungs. Moreover, we observe that the lesions of various sizes appear randomly in the lungs, resulting in only a portion of the slices containing visible disease characterizations. The spatial correlations of different dimensions and the inter-slice information will be entangled by a 3D convolution operator when using 3D CNN to directly classify CT images. To refine the hidden features, Hu et al. \cite{25} proposed the channel-wise attention module to enhance channel independence and thereby improve the performance of the networks. But this module has limitations in our task, due to the sparse distribution of lesion features at the depth level. Motivated by this observation, we design a complementary mechanism called depth-wise attention module for 3D CNN to recalibrate the depth-level features, which can make the network more sensitive to the important regions of the images. By integrating DA and CA modules, we construct the dual-attention module used in DARNet.

\subsubsection{Channel-wise Attention Module}
We implement the 3D version of CA module based on the origin idea in \cite{25}, as shown in Fig \ref{ca}. Firstly, the input features are squeezed by a GAP layer. Considering the input feature map $\mathbf{F_{in}}\in\mathbb{R}^{C\times D\times H \times W}$ and  $\mathbf{F_{in}} = [\mathbf{f}_{1},\mathbf{f}_{2},...,\mathbf{f}_{C}]$, where $C$, $D$, $H$, and $W$ are the input channels, depth, height, and width, respectively, and $\mathbf{f}_{i} \in \mathbb{R}^{D \times H\times W}$. The output of the GAP represented by  $\mathbf{Z} \in \mathbb{R}^{C \times 1 \times 1 \times 1}$ with its element

\begin{equation}
	z_{i} = \frac{1}{D \times H\times W} \sum\limits_{d=1}^{D}\sum\limits_{h=1}^{H}\sum\limits_{w=1}^{W}\mathbf{f}_{i}(d,h,w).
\end{equation}

Above operation embeds the global spatial information in vector $\mathbf{Z}$. This vector is transformed to the weight vector $\mathbf{\hat Z} = \sigma(\mathbf{W_2}(\xi(\mathbf{W_1}\mathbf{Z})))$, with $\mathbf{W_1}\in\mathbb{R}^{\frac{C}{r}\times C}$, $\mathbf{W_2}\in\mathbb{R}^{C\times\frac{C}{r}}$ being the weights of two fully-connected layers, the ReLU function $\xi(\cdot)$ and the sigmoid function $\sigma(\cdot)$. The parameter $r$ refers to the reduction ratio and is set to 16 in this study. The recalibrated output vector is 

\begin{equation}
	\mathbf{F_{out}} = [\hat z_{1}\mathbf{f}_{1},\hat z_{2}\mathbf{f}_{2},...,\hat z_{i}\mathbf{f}_{i},...,\hat z_{C}\mathbf{f}_{C}].
\end{equation}

Each element in $\mathbf{\hat Z}$ indicates the importance of the corresponding channel and is used to dynamically amplify or suppress the input response. In this way, the CA module can enhance the important features and ignore the irrelevant ones. However, it is limited to directly extend this module in 3D neural networks to classify CT images. Due to the sparse distribution of lesions, the information between slices varies greatly. The performance improvement achieved by differentiating channel-level features alone is not very significant. Therefore, we design the DA module to make up for this defect.
 
\subsubsection{Depth-wise Attention Module}

For DA module, as same as CA module, the spatial information is aggregated first along the depth axis by GAP layer, as shown in Fig. \ref{da}. Considering the input feature map $\mathbf{U_{in} }\in\mathbb{R}^{C\times D\times H \times W}$ and  $\mathbf{U_{in}} = [\mathbf{u}^{1,1},\mathbf{u}^{1,2},...,\mathbf{u}^{i,j},...,\mathbf{u}^{C,D}]$, and $\mathbf{u}^{i,j} \in \mathbb{R}^{H\times W}$. The output of the GAP represented by  $\mathbf{T} \in \mathbb{R}^{C \times D \times 1 \times 1}$ with its element

\begin{equation}
	t^{i,j} = \frac{1}{H\times W} \sum\limits_{h=1}^{H}\sum\limits_{w=1}^{W}\mathbf{u}^{i,j}(h,w).
\end{equation}

Then, a gating mechanism is designed to the learn non-linear and non-mutually-exclusive relationships in the depth dimension. The gating mechanism is parameterized by two fully-connected layers and two non-linearity activation functions. The output is $\mathbf{\hat T} = \sigma(\mathbf{W_2}(\xi(\mathbf{W_1}\mathbf{T})))$, with $\mathbf{W_1}\in\mathbb{R}^{\frac{CD}{r}\times CD}$, $\mathbf{W_2}\in\mathbb{R}^{CD\times\frac{CD}{r}}$ being the weights of two fully-connected layers. The parameter $r$ here is equal to the number of input channels. Finally, the resultant tensor is used to refine $\mathbf{U_{in}}$ to

\begin{equation}
	\mathbf{U_{out}} = [\hat t^{1,1}\mathbf{u}^{1,1},\hat t^{1,2}\mathbf{u}^{1,2},...,\hat t^{i,j}\mathbf{u}^{i,j},...,\hat t^{C,D}\mathbf{u}^{C,D}].
\end{equation}

The DA module recalibrates the depth-level features by adaptively assigning weights, which can make the network more focused on the important regions distributed sparsely along the depth dimension. This module makes up for the deficiency of the CA module. Then, we develop the dual-attention module of DARNet based on the serial combination of the two, which can refine the learned features at different levels.

\section{Experiments}

We conduct experiments on a public dataset provided by the China Consortium of Chest CT Image Investigation (CC-CCII\footnote{http://ncov-ai.big.ac.cn/download?lang=en}) \cite{21} to evaluate our method. In this section, the construction of the dataset used and implementation details are described first. Then, we compare different networks in terms of the diagnostic performance, and perform ablation studies to validate the effectiveness of the proposed dual-attention module in improving the performance. Finally, class activation mapping (CAM) \cite{30} is employed to visualize the discriminative regions of these networks in diagnosing COVID-19, which can help to explore the interpretability of different methods.

\subsection{Dataset and Metrics}

\begin{table}[htbp]
	\caption{The statistics and division of the experimental dataset. The dataset is divided into the training and test sets.}
	\setlength{\tabcolsep}{1mm}
	\renewcommand{\arraystretch}{1.4}
	\centering
	\begin{tabular}{C{2.5cm}|C{1.8cm}|C{1.8cm}|C{1.8cm}}
		\hline
		\#images (patients) & Training set & Test set & In total\\ \hline
		
		COVID-19 & 1,245 (764) & 299 (165)   &1,544 (929) \\ \hline
		CP & 1,137 (789)     & 419 (175)        & 1,556 (964)     \\  \hline
		healthy controls  & 856 (732)      & 222 (117)       & 1,078 (849)      \\  \hline
		In total & 3,238 (2,285)   & 940 (457)     & 4,178 (2,742)   \\ \hline
	\end{tabular}
	\label{dataset}
\end{table}

In this paper, we evaluate our proposed method on a large publicly available CT dataset provided by CC-CCII. The CT dataset contains a total of 4,178 chest CT images from 2,742 patients, including 1,544 CT images from 929 COVID-19 patients, 1,556 CT images from 964 CP patients, and 1,078 CT images from 849 healthy controls. As shown in Table \ref{dataset}, we separate the dataset into two parts. The first part (Training set) is used for training, which includes 1,245 COVID-19 images, 1,137 CP images, and 856 images of healthy controls. The second part (Test set) serves for independent testing, including 299 COVID-19 images, 419 CP image, and 222 images of healthy controls. In particular, the split is done on patient level, which means the images of same subject are kept in the same set of training or testing.

In the training stage, the training set is randomly divided into five folds on patient level for cross-validation. For evaluating, we use five different classification metrics, including the area under the receiver operating characteristic curve (AUC), accuracy, sensitivity, specificity, and F1-score, to evaluate the performance of different networks. The mathematical expressions of accuracy, sensitivity, and specificity are shown below. 

\begin{equation}
	Accuracy = \frac{TP+TN}{TP+TN+FP+FN}.
\end{equation}

\begin{equation}
	Sensitivity = \frac{TP}{TP+FN}.
\end{equation}

\begin{equation}
	Specificity = \frac{TN}{TN+FP}.
\end{equation}

True positive, true negative, false positive, and false negative are denoted by TP, TN, FP, and FN respectively.

\begin{table*}[]
	\caption{The performance comparison between different methods of identifying COVID-19 on the dataset provided by CC-CCII. For the results on the independent test set of DARNet, we show the mean$\pm$std (standard deviation) scores of five trained models of each training-validation fold. Larger values indicate better performance, and - denotes no relevant data.}
	\setlength{\tabcolsep}{1mm}
	\centering
	\renewcommand{\arraystretch}{1.4}
	\begin{tabular}{C{3cm}|C{2.2cm}|C{2.2cm}|C{2.2cm}|C{2.2cm}|C{2.2cm}}
		\hline
		Method      & AUC    & Accuracy (\%) & Sensitivity (\%)  & Specificity (\%)  & F1-score (\%)   \\ [2pt]\hline
		\cite{21}  & 0.9797   & 92.49  & 94.93  & 91.13  & -      \\ [2pt]
		\cite{36}  & -   & 88.69  & 88.08  & -  & 89.26  \\ [2pt]
		\cite{39}  & -        & \textbf{93.57}  & 94.21  & 93.93  & 91.74  \\ [2pt]
		\cite{35}  & 0.9212   & -      & 77.99  & 93.55  & -      \\ [2pt]
		\textbf{DARNet}  & \textbf{0.9950$\pm$0.0020}   & 93.28$\pm$0.85   & \textbf{96.86$\pm$0.81}   & \textbf{97.19$\pm$1.08}   & \textbf{95.49$\pm$1.15} \\ [2pt]\hline
	\end{tabular}
	\label{cdm}
\end{table*}

\subsection{Implementation Details}
Pytorch is adopted to implement our proposed method. For training the networks, we use Adam optimizer \cite{32} to minimize the cross-entropy loss with an initial learning rate of $10^{-3}$. The convolutional layer weights are initialized by the Kaiming Normalization \cite{33} and the biases are set to 0. Besides, we apply the multi-step decay strategy to control the change of the learning rate during training. The learning rate is reduced every 30 epochs with a decay factor of 0.1. All the models are trained from scratch using 2 NVIDIA Tesla P40 graphic processing units. Given the limitation of GPU memory, the batch size is set to 8 and the size of all images is fixed to  $64\times224\times224$ by under-sampling or up-sampling. In each fold, the model is evaluated on the validation set at the end of each training epoch, and finally the best model within 80 epochs is evaluated on the independent test set. To alleviate the overfitting problem, we conduct online data augmentation including random flipping, rotation, translation, and scaling. The codes used in the experiments is available\footnote{https://github.com/shijun18/COVID-19\_CLS}.

\subsection{Overall Performance}

We compare the performance of DARNet with four existing methods. For a fair comparison, the test sets used by these methods are also from the same CT dataset provided by CC-CCII, and we directly quote the results reported in related papers. As shown in Table \ref{cdm}, we can see that DARNet achieves the best performance on four indicators with sensitivity of 96.86\%, specificity of 97.19\%, F1-score of 95.49\%, and AUC of 0.995. As for the accuracy, the performance of DARNet is a little bit lower than that of \cite{39}. 

In particular, \cite{39} proposed an ensemble learning method using multiple classifiers to make the diagnostic decision. Although this method has high accuracy, it is also demanding on the classifier design and integration strategy. \cite{36} provides a benchmark for COVID-19 detection using deep learning models. The benchmark tests multiple models and we select the best performing one for comparison. According to its results, we observe that it is limited to directly transfer 2D neural networks to classify 3D CT images. The main reason is that this method ignores the rich spatial information preserved in the 3D structure. Moreover, \cite{21} and \cite{35} are segmentation-based methods, which highly rely on accurate segmentation of the lesions. However, these multi-stage frameworks often suffer from error propagation. For example, the incorrect segmentation results can directly make a negative impact on subsequent tasks. In contrast, DARNet is an end-to-end model that can avoid this problem. Besides, the proposed dual-attention module can effectively improve the feature extraction ability of the model, which helps to obtain higher classification performance than the naive CNN-based methods. The results in Table \ref{cdm} prove the superiority of DARNet in identifying COVID-19 from CP and healthy people.

\subsection{Ablation Study}

The overall experiments have proved the superiority of DARNet. However, which module plays a more important role in performance improvement is still unclear. Therefore, we conduct an ablation study to validate the effectiveness of each module, including CA, DA, and the dual-attention modules. Table \ref{two_and_three} quantitatively compares the performance of different networks on the independent test set. For COVID-19 versus the other two classes (CP and healthy controls), DARNet achieves the highest AUC, accuracy, sensitivity, and F1-score. Meanwhile, DARNet obtains the best results on all performance indicators for the three-way classification. 

\begin{table}[H]
	\caption{Parameter comparison of ablation study.}
	\setlength{\tabcolsep}{1mm}
	\centering
	\renewcommand{\arraystretch}{1.4}
	\begin{tabular}{C{4cm}|C{3cm}}
		\hline
		Method      & Parameters (M)   \\ [3pt]\hline
		DARNet (w/o DA and CA)    & 33.15 \\ [3pt] 
		DARNet (w/o DA) & 33.24 \\ [3pt] 
		DARNet (w/o CA) & 35.15  \\ [3pt] 
		DARNet & 35.24  \\ [3pt]\hline
	\end{tabular}
	\label{param}
\end{table}

\begin{table*}[]
	\caption{For evaluating the effectiveness of the proposed module, we conduct an ablation study (w/o denotes without) on the independent test set. The result of accuracy indicator is the macro average of all categories.}
	\setlength{\tabcolsep}{1mm}
	\centering
	\renewcommand{\arraystretch}{1.4}
	\begin{tabular}{C{3.2cm}|C{2.16cm}|C{2.16cm}|C{2.16cm}|C{2.16cm}|C{2.16cm}}
		\hline
		Method      & AUC    & Accuracy (\%) & Sensitivity (\%)  & Specificity (\%)  & F1-score (\%)   \\ [2pt]\hline
		\multicolumn{6}{l}{Two classes: COVID-19 and non-COVID-19 (CP and healthy) }\\  [2pt] \hline
		DARNet (w/o DA and CA)     & 0.9938$\pm$0.0024  & 91.32$\pm$1.87  & 95.52$\pm$1.07  & 97.15$\pm$1.28 & 94.86$\pm$1.21 \\ [2pt] 
		DARNet (w/o DA)    & 0.9926$\pm$0.0024  & 92.13$\pm$1.90  & 94.65$\pm$1.86  & \textbf{97.88$\pm$1.28} & 95.04$\pm$0.49\\ [2pt] 
		DARNet (w/o CA) & 0.9923$\pm$0.0024 & 92.55$\pm$0.51  & 95.25$\pm$0.73  & 97.72$\pm$1.15 & 95.20$\pm$0.89 \\ [2pt] 
		\textbf{DARNet}  & \textbf{0.9950$\pm$0.0020}   & \textbf{93.28$\pm$0.85}   & \textbf{96.86$\pm$0.81}   & 97.19$\pm$1.08   & \textbf{95.49$\pm$1.15} \\ [2pt]\hline
		\multicolumn{6}{l}{Three classes: COVID-19, CP and healthy}\\  [2pt] \hline
		DARNet (w/o DA and CA)     & 0.9850$\pm$0.0041  & 91.32$\pm$1.87  & 90.34$\pm$3.85  & 95.32$\pm$1.88 & 90.74$\pm$2.44 \\ [2pt] 
		DARNet (w/o DA)    & 0.9854$\pm$0.0047  & 92.13$\pm$1.90  & 91.70$\pm$3.58  & 95.95$\pm$1.95 & 91.63$\pm$2.44\\ [2pt] 
		DARNet (w/o CA) & 0.9832$\pm$0.0041 & 92.55$\pm$0.51  & 91.96$\pm$1.55  & 96.14$\pm$1.00 & 92.05$\pm$0.81 \\ [2pt] 
		\textbf{DARNet}  & \textbf{0.9879$\pm$0.0028}   & \textbf{93.28$\pm$0.85}   & \textbf{93.03$\pm$1.85}  & \textbf{96.56$\pm$0.99}   & \textbf{92.92$\pm$1.09} \\ [2pt]\hline
		
	\end{tabular}
	\label{two_and_three}
\end{table*}

\begin{figure*}[]
	\centering
	\includegraphics[width=1.75\columnwidth]{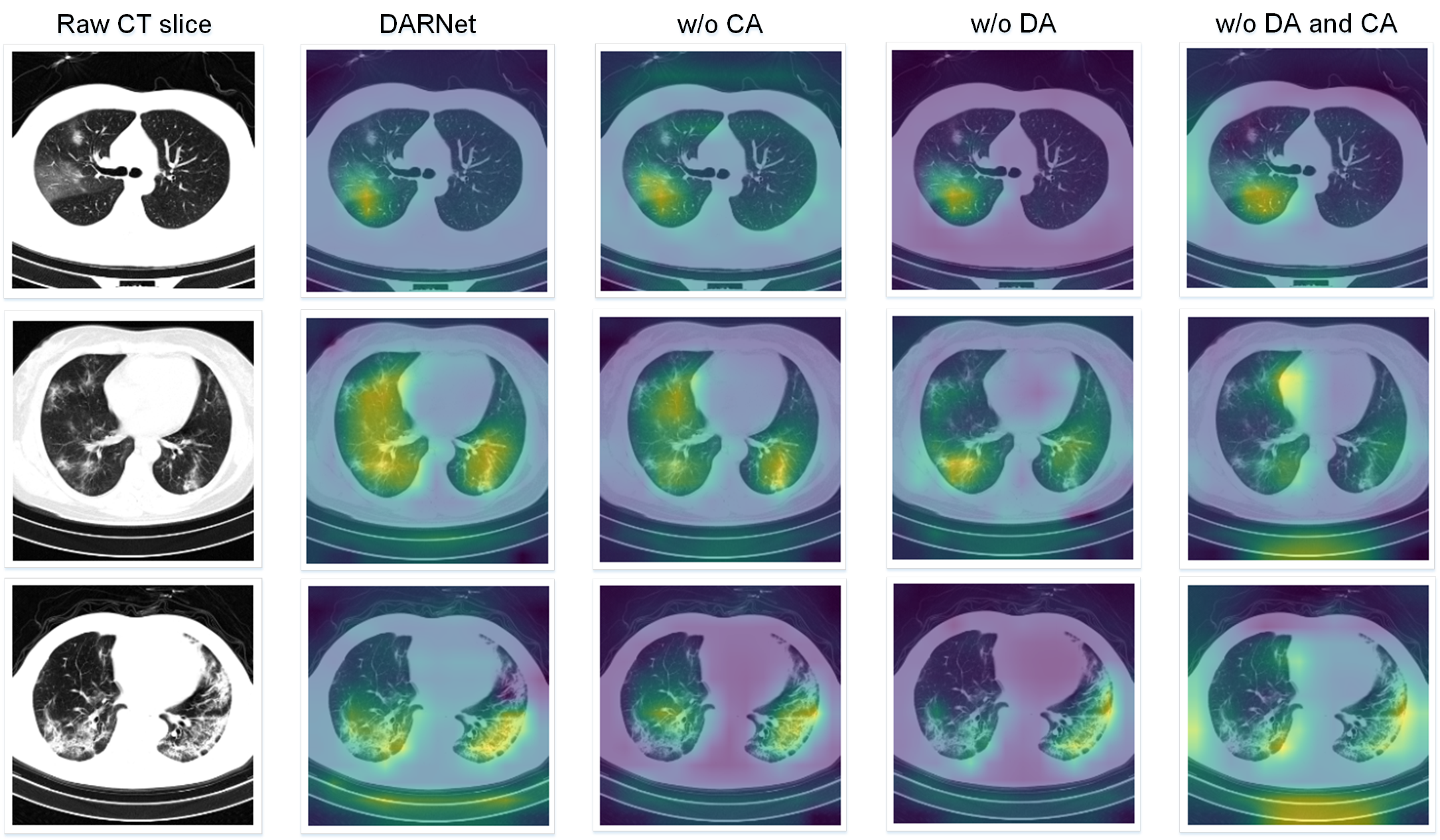} 
	\caption{Visualization results of different methods on three COVID-19 cases with varying degrees of infection. From top to down, there are mild, moderate, and severe cases, respectively. The discriminative regions of different networks are highlighted.}
	\label{va}
\end{figure*}

The results of the ablation experiments reveal the importance of each part. According to the results, we can observe varying degrees of model performance decline. Among all of them, the dual-attention module has the biggest impact on the model performance. By applying the dual-attention module, DARNet has a significant improvement on all performance indicators, while the parameter is only increased by about $6.4\%$ as shown in Table \ref{param}. Moreover, removing CA or DA module will also have a negative impact on network performance. These observations further prove the effectiveness of the dual-attention module.

\subsection{Visualization Analysis}

To further explore the interpretability of DARNet, we employ CAM \cite{30} to visualize the discriminative regions of different networks in diagnosing COVID-19. Fig. \ref{va} shows the visualization results on three COVID-19 cases with different degrees (mild, moderate, and severe) of infection, highlighting the regions that the network focuses on when making decisions. We observe that DARNet can accurately locate lung lesions that vary greatly in size and distribution.

However, after removing CA or DA module, the positioning ability of the network has declined significantly. For instance, for the severe COVID-19 case in Fig. \ref{va}, we can see diffuse lesions in both lungs, consolidation of the lower lobe of the left lung. When we remove the CA and DA modules in turn, the highlighted area in the right lung gradually shrinks. Especially, the network without these two modules has very low sensitivity to the lesions, and may even be disturbed by the information outside the lung area. The above results demonstrate that DA and CA modules can enhance the learned features to ensure that the decisions made by the network depend mainly on the infection regions to a certain extent, rather than the irrelevant parts of the images. More importantly, the results also show that DARNet has better interpretability and reliability in diagnosing COVID-19.

\section{Conclusion}

In this work, we proposed a dual-attention residual network that can realize the automatic and accurate diagnosis of COVID-19 using 3D chest CT images. In our method, we constructed the dual-attention module by combining CA and DA modules to refine the hidden features by adaptively assigning weights during training. This module can effectively improve the classification performance and interpretability of 3D ResNet, while only slightly increasing the computational complexity. We evaluated our method on a large public CT dataset, achieving state-of-the-art results. To further understand the decision of the proposed method, we showed the visual evidence to reveal the discriminative regions used in the model for diagnosis. In future work, we will further investigate the generalization capability of the proposed method. Besides, more work is still devoted to analyzing the relationship between these discriminative regions and the image findings.

\bibliographystyle{IEEEtran}
\bibliography{mybib}

\begin{thebibliography}{10}
\providecommand{\url}[1]{#1}
\csname url@samestyle\endcsname
\providecommand{\newblock}{\relax}
\providecommand{\bibinfo}[2]{#2}
\providecommand{\BIBentrySTDinterwordspacing}{\spaceskip=0pt\relax}
\providecommand{\BIBentryALTinterwordstretchfactor}{4}
\providecommand{\BIBentryALTinterwordspacing}{\spaceskip=\fontdimen2\font plus
\BIBentryALTinterwordstretchfactor\fontdimen3\font minus
  \fontdimen4\font\relax}
\providecommand{\BIBforeignlanguage}[2]{{%
\expandafter\ifx\csname l@#1\endcsname\relax
\typeout{** WARNING: IEEEtran.bst: No hyphenation pattern has been}%
\typeout{** loaded for the language `#1'. Using the pattern for}%
\typeout{** the default language instead.}%
\else
\language=\csname l@#1\endcsname
\fi
#2}}
\providecommand{\BIBdecl}{\relax}
\BIBdecl

\bibitem{1}
J.~F.-W. Chan, S.~Yuan, K.-H. Kok, and et~al., ``A familial cluster of
  pneumonia associated with the 2019 novel coronavirus indicating
  person-to-person transmission: a study of a family cluster,'' \emph{The
  Lancet}, vol. 395, no. 10223, pp. 514--523, Feb 2020.

\bibitem{2}
E.~Dong, H.~Du, and L.~Gardner, ``An interactive web-based dashboard to track
  covid-19 in real time,'' \emph{The Lancet Infectious Diseases}, vol.~20,
  no.~5, pp. 533--534, May 2020.

\bibitem{3}
Z.~Y. Zu, M.~{Di Jiang}, P.~P. Xu, W.~Chen, Q.~Q. Ni, G.~M. Lu, and L.~J.
  Zhang, ``{Coronavirus Disease 2019 (COVID-19): A Perspective from China},''
  \emph{Radiology}, vol. 296, no.~2, pp. E15--E25, 2020.

\bibitem{6}
S.~Kadry, V.~Rajinikanth, S.~Rho, N.~S.~M. Raja, V.~S. Rao, and K.~P. Thanaraj,
  ``{Development of a Machine-Learning System to Classify Lung CT Scan Images
  into Normal/COVID-19 Class},'' \emph{arXiv preprint arXiv: 2004.13122}, 2020.

\bibitem{7}
T.~Ai, Z.~Yang, H.~Hou, and et~al., ``{Correlation of Chest CT and RT-PCR
  Testing for Coronavirus Disease 2019 (COVID-19) in China: A Report of 1014
  Cases},'' \emph{Radiology}, vol. 296, no.~2, pp. E32--E40, 2020.

\bibitem{8}
A.~L. Phelan, R.~Katz, and L.~O. Gostin, ``{The Novel Coronavirus Originating
  in Wuhan, China: Challenges for Global Health Governance},'' \emph{JAMA},
  vol. 323, no.~8, pp. 709--710, 02 2020.

\bibitem{9}
H.~Nishiura, S.-M. Jung, N.~M. Linton, R.~Kinoshita, Y.~Yang, and et~al.,
  ``\BIBforeignlanguage{eng}{The extent of transmission of novel coronavirus in
  wuhan, china, 2020},'' \emph{\BIBforeignlanguage{eng}{Journal of clinical
  medicine}}, vol.~9, no.~2, p. 330, Jan 2020.

\bibitem{10}
X.~Mei, H.~C. Lee, and et~al., ``{Artificial intelligence–enabled rapid
  diagnosis of patients with COVID-19},'' \emph{Nature Medicine}, vol.~26,
  no.~8, pp. 1224--1228, 2020.

\bibitem{11}
C.~Butt, J.~Gill, D.~Chun, and B.~A. Babu, ``{Deep learning system to screen
  coronavirus disease 2019 pneumonia},'' \emph{Applied Intelligence}, 2020.

\bibitem{20}
A.~M. Hasan, M.~M. Al-Jawad, and et~al., ``{Classification of Covid-19
  coronavirus, pneumonia and healthy lungs in CT scans using Q-deformed entropy
  and deep learning features},'' \emph{Entropy}, vol.~22, no.~5, 2020.

\bibitem{21}
K.~Zhang, X.~Liu, J.~Shen, et~al., J.~He, and et~al., ``Clinically applicable
  ai system for accurate diagnosis, quantitative measurements, and prognosis of
  covid-19 pneumonia using computed tomography,'' \emph{Cell}, vol. 181, no.~6,
  pp. 1423 -- 1433.e11, 2020.

\bibitem{22}
X.~Ouyang, J.~Huo, L.~Xia, and et~al., ``{Dual-Sampling Attention Network for
  Diagnosis of COVID-19 from Community Acquired Pneumonia},'' \emph{IEEE
  Transactions on Medical Imaging}, vol.~39, no.~8, pp. 2595--2605, 2020.

\bibitem{23}
L.~Li, L.~Qin, Z.~Xu, and et~al., ``\BIBforeignlanguage{eng}{Using artificial
  intelligence to detect covid-19 and community-acquired pneumonia based on
  pulmonary ct: Evaluation of the diagnostic accuracy},''
  \emph{\BIBforeignlanguage{eng}{Radiology}}, vol. 296, no.~2, pp. E65--E71,
  Aug 2020.

\bibitem{36}
X.~He, S.~Wang, X.~Chu, S.~Shi, J.~Tang, X.~Liu, C.~Yan, J.~Zhang, and G.~Ding,
  ``Automated model design and benchmarking of deep learning models for
  covid-19 detection with chest ct scans,'' \emph{Proceedings of the AAAI
  Conference on Artificial Intelligence}, vol.~35, no.~6, pp. 4821--4829, May
  2021.

\bibitem{39}
X.~Li, W.~Tan, P.~Liu, Q.~Zhou, and J.~Yang, ``{Classification of COVID-19
  Chest CT Images Based on Ensemble Deep Learning},'' \emph{Journal of
  Healthcare Engineering}, vol. 2021, 2021.

\bibitem{13}
K.~He, X.~Zhang, S.~Ren, and J.~Sun, ``Deep residual learning for image
  recognition,'' in \emph{Proceedings of the IEEE conference on computer vision
  and pattern recognition}, 2016, pp. 770--778.

\bibitem{25}
J.~Hu, L.~Shen, and G.~Sun, ``Squeeze-and-excitation networks,'' in
  \emph{Proceedings of the IEEE conference on computer vision and pattern
  recognition}, 2018, pp. 7132--7141.

\bibitem{17}
G.~Litjens, T.~Kooi, B.~E. Bejnordi, and et~al., ``A survey on deep learning in
  medical image analysis,'' \emph{Medical Image Analysis}, vol.~42, pp. 60 --
  88, 2017.

\bibitem{19}
J.~Zhang, Y.~Xie, Z.~Liao, G.~Pang, J.~Verjans, W.~Li, Z.~Sun, J.~He, Y.~Li,
  C.~Shen, and Y.~Xia, ``{Viral Pneumonia Screening on Chest X-ray Images Using
  Confidence-Aware Anomaly Detection},'' \emph{arXiv preprint arXiv:
  2003.11988}, 2020.

\bibitem{40}
X.~Li, C.~Li, and D.~Zhu, ``{COVID-MobileXpert: On-Device COVID-19 Patient
  Triage and Follow-up using Chest X-rays},'' \emph{Proceedings - 2020 IEEE
  International Conference on Bioinformatics and Biomedicine, BIBM 2020}, pp.
  1063--1067, 2020.

\bibitem{41}
M.~R. Karim, T.~Dohmen, M.~Cochez, O.~Beyan, D.~Rebholz-Schuhmann, and
  S.~Decker, ``{DeepCOVIDExplainer: Explainable COVID-19 Diagnosis from Chest
  X-ray Images},'' \emph{Proceedings - 2020 IEEE International Conference on
  Bioinformatics and Biomedicine, BIBM 2020}, pp. 1034--1037, 2020.

\bibitem{27}
Y.~Cao, J.~Xu, S.~Lin, F.~Wei, and H.~Hu, ``{GCNet: Non-local networks meet
  squeeze-excitation networks and beyond},'' \emph{Proceedings - 2019
  International Conference on Computer Vision Workshop, ICCVW 2019}, pp.
  1971--1980, 2019.

\bibitem{28}
S.~Gao, M.-M. Cheng, K.~Zhao, X.-Y. Zhang, M.-H. Yang, and P.~H. Torr,
  ``{Res2Net: A New Multi-scale Backbone Architecture},'' \emph{IEEE
  Transactions on Pattern Analysis and Machine Intelligence}, vol. 8828, no.~c,
  pp. 1--1, 2019.

\bibitem{29}
Q.~Zhao, T.~Sheng, Y.~Wang, Z.~Tang, Y.~Chen, L.~Cai, and H.~Ling, ``{M2Det: A
  Single-Shot Object Detector Based on Multi-Level Feature Pyramid Network},''
  \emph{Proceedings of the AAAI Conference on Artificial Intelligence},
  vol.~33, pp. 9259--9266, 2019.

\bibitem{42}
A.~G. Roy, N.~Navab, and C.~Wachinger, ``{Recalibrating Fully Convolutional
  Networks With Spatial and Channel 'Squeeze and Excitation' Blocks},''
  \emph{IEEE Transactions on Medical Imaging}, vol.~38, no.~2, pp. 540--549,
  2019.

\bibitem{43}
J.~Fu, J.~Liu, H.~Tian, Y.~Li, Y.~Bao, Z.~Fang, and H.~Lu, ``{Dual attention
  network for scene segmentation},'' \emph{Proceedings of the IEEE Computer
  Society Conference on Computer Vision and Pattern Recognition}, vol.
  2019-June, pp. 3141--3149, 2019.

\bibitem{30}
B.~{Zhou}, A.~{Khosla}, A.~{Lapedriza}, A.~{Oliva}, and A.~{Torralba},
  ``Learning deep features for discriminative localization,'' in \emph{2016
  IEEE Conference on Computer Vision and Pattern Recognition (CVPR)}, 2016, pp.
  2921--2929.

\bibitem{35}
C.~Jin, W.~Chen, Y.~Cao, et~al. Jie, H.~Shi, and J.~Feng, ``{Development and
  evaluation of an artificial intelligence system for COVID-19 diagnosis},''
  \emph{Nature Communications}, vol.~11, no.~1, 2020.

\bibitem{32}
D.~P. Kingma and J.~Ba, ``Adam: A method for stochastic optimization,''
  \emph{arXiv preprint arXiv: 1412.6980}, 2014.

\bibitem{33}
K.~He, X.~Zhang, S.~Ren, and J.~Sun, ``Delving deep into rectifiers: Surpassing
  human-level performance on imagenet classification,'' in \emph{Proceedings of
  the IEEE international conference on computer vision}, 2015, pp. 1026--1034.

\end{thebibliography}

\end{document}